# Enhanced interface perpendicular magnetic anisotropy in Ta|CoFeB|MgO using nitrogen doped Ta underlayers

Jaivardhan Sinha[1], Masamitsu Hayashi[1*], Andrew J. Kellock[2], Shunsuke Fukami[3], Michihiko Yamanouchi[3,4], Hideo Sato[3], Shoji Ikeda[3,4], Seiji Mitani[1], See-hun Yang[2], Stuart S. P. Parkin[2] and Hideo Ohno[3,4,5]

[1]*National Institute for Materials Science, Tsukuba 305-0047, Japan*
[2]*IBM Almaden Research Center, San Jose, CA 95120, USA*
[3] *Center for Spintronics Integrated Systems, Tohoku University, Sendai 980-8577, Japan*
[4]*Laboratory for Nanoelectronics and Spintronics, Research Institute of Electrical Communication, Tohoku University, Sendai 980-8577, Japan*
[5]*WPI Advanced Institute for Materials Research, Tohoku University, Sendai 980-8577, Japan*

We show that the magnetic characteristics of Ta|CoFeB|MgO magnetic heterostructures are strongly influenced by doping the Ta underlayer with nitrogen. In particular, the saturation magnetization drops upon doping the Ta underlayer, suggesting that the doped underlayer acts as a boron diffusion barrier. In addition, the thickness of the magnetic dead layer decreases with increasing nitrogen doping. Surprisingly, the interface magnetic anisotropy increases to ~1.8 erg/cm$^2$ when an optimum amount of nitrogen is introduced into the Ta underlayer. These results show that nitrogen doped Ta serves as a good underlayer for Spintronics applications including magnetic tunnel junctions and domain wall devices.

*Email: hayashi.masamitsu@nims.go.jp

Perpendicular magnetic anisotropy (PMA) is one of the key material parameters in modern spintronics devices.[1] The threshold current needed to switch the direction of magnetic moments in magnetic tunnel junctions and the current required to move magnetic domain walls in magnetic nanowires can be significantly reduced by introducing perpendicularly magnetized materials into the system.[2-5] For device application purposes, the magnetic layer needs to be thin enough such that current induced torques, including conventional spin transfer torques[6, 7] and more recently discovered spin orbit torques,[8-10] can act on the magnetic moments to switch the magnetization and/or move domain walls. An attractive step towards building suitable materials systems for such applications is to make use of the interface magnetic anisotropy, which plays an important role in thin film heterostructures, in particular, in metallic multilayers.[11, 12] Interface PMA allows the system to maintain its magnetic anisotropy down to small thicknesses, whereas the bulk anisotropy, including crystalline and magneto-elastic anisotropies, are lowered as the thickness is reduced.[13]

Recently, it has been reported that significant interface PMA exists in CoFeB|MgO,[2] a system that is at the heart of advanced magnetic tunnel junctions. As the CoFeB thickness is reduced, the magnetic easy axis changes from lying within the film plane to lying along the film normal owing to the large PMA at the CoFeB|MgO interface. Recent theoretical calculations suggest that the large PMA at this interface is due to an electronic effect.[14] On the other hand, it has been experimentally shown that the underlayer immediately adjacent to the opposite side of the CoFeB layer also plays a critical role in determining the magnitude of PMA.[3, 15] Up to date, Ta (or, more recently, Hf[15]) is considered to be the optimal choice for introducing large PMA for CoFeB|MgO.

However, Ta is known for creating a magnetic dead layer when placed next to a magnetic layer.[16-18] Since the thickness of the CoFeB layer needs to be of the order of ~1 nm to utilize moderate interface PMA, the presence of a magnetic dead layer, if any, will be of a particular concern for thermal stability issues. It is also important to understand the Ta|CoFeB interface for devices that take advantage of spin orbit torques[9, 10, 19] (for example, spin torque generated from the spin Hall effect in the Ta layer) for which transparent interfaces are required for passing spin currents across the bilayer.

Here we show the magnetic properties of CoFeB|MgO heterostructures with Ta and doped-Ta underlayers. First, we find that the saturation magnetization changes below and above a critical CoFeB thickness of ~2 nm. Using the thinner regime for characterization, a non-zero magnetic dead layer is found, likely at the interface of Ta|CoFeB. We use nitrogen doping to reduce the dead layer thickness, a process that has been proved to work in other systems.[20] We find that a small amount of $N_2$ introduced into the Argon sputter gas (~2 molecular %) is sufficient to reduce the dead layer thickness to near zero. Interestingly, the interface PMA reaches a maximum value of ~1.8 erg/cm$^2$ at a $N_2$ gas concentration of ~1%, which is higher than that of the non-doped Ta underlayer stack. In addition, in this thin CoFeB regime, we find that there is a considerable volume contribution to the PMA which is not negligible.

All films are deposited at room temperature using magnetron sputtering on 10 x 10 mm$^2$ square Si (001) substrates, coated with 100 nm thick $SiO_2$. The film stack is the following: Substrate|d Ta(N)|t $Co_{20}Fe_{60}B_{20}$|2 MgO|1 Ta (units in nanometers). Here d and t correspond to the nominal thicknesses of the Ta(N) and CoFeB layers, respectively. The $Ta_{1-X}N_X$ (abbreviated as "TaN" hereafter) underlayer is formed by mixing $N_2$ gas into the Ar gas atmosphere during the sputtering of Ta. We define a quantity Q, the ratio between the $N_2$ ($S_{N2}$) and the Ar ($S_{Ar}$) gas flow, measured using a mass flow meter attached to each gas line, to represent the $N_2$ concentration ($Q=S_{N2}/(S_{Ar}+S_{N2})\times 100$) in the sputtering gas atmosphere. Q is varied from 0, corresponding to pure Ta underlayer, to ~9%. All films are post-annealed at 300 ˚C for one hour in a vacuum chamber: no magnetic field is applied during the annealing.

Saturation magnetization ($M_S$) and magnetic anisotropy energy ($K_{EFF}$) are measured at room temperature using vibrating sample magnetometry (VSM). Each film is deposited on the substrate through a shadow mask which defines the area (A) of the film. $K_{EFF}$ is estimated from the areal difference between the out of plane and in-plane magnetization hysteresis loops. Positive $K_{EFF}$ corresponds to the magnetic easy axis directed along the film normal. Rutherford Backscattering Spectroscopy (RBS) is used to study the composition of the TaN films. We use a single layer structure (no capping layer) for the composition analysis; the thicknesses of these films are ~20-50 nm. Four point probe resistivity measurements are performed for these films prior to the RBS analysis.

Figure 1(a) shows the film composition (N/Ta ratio, $x/(1-x)$) of the single layer $Ta_{1-X}N_X$ films plotted as a function of the $N_2$ to Ar gas flow ratio Q. The horizontal dashed lines indicate the stable $Ta_{1-X}N_X$ compounds as a reference.[21] The nitrogen concentration abruptly increases upon introduction of the $N_2$ gas into the sputtering gas atmosphere. The Q dependence of the TaN composition depends on the Ar+$N_2$ gas pressure (here it is fixed to ~1.1 Pa) and specific details of the sputtering chamber, in particular, on how each gas flows inside the chamber and approaches the substrate.[21] In contrast to the composition, the resistivity of the films (Fig. 1(a) inset) scales with the gas flow ratio for small Q (Q≤2.5%). The relationship between the N/Ta ratio and the resistivity is similar to what has been reported previously[21] (see supplementary material[22]).

The CoFeB thickness (t) dependence of the magnetic moment per unit area (M/A) measured at room temperature (circles) is shown in Fig. 1(b) for the Ta underlayer films (Q=0). The data is fitted with a linear function to obtain the *average* saturation magnetization ($\tilde{M}_S$). Here we find that the slope of M/A vs. t changes at t~2.2 nm, suggesting a change in $\tilde{M}_S$ at a critical CoFeB thickness (defined as $t_C$ hereafter). It is typically assumed that $\tilde{M}_S$ gradually decreases from its bulk value below a certain magnetic layer thickness due to a reduction in the Curie temperature. We have studied the temperature dependence of M/A and found that the slope change also takes place at temperatures as low as ~10 K, as shown by the square symbols in Fig. 1(b), indicating that the Curie temperature variation is not the main cause of the slope change. Structural difference, such as amorphous CoFeB for the thicker regime and (partially) textured polycrystalline BCC CoFeB for the thinner side,[23] may explain the change in the magnetization below and above $t_C$, although we do not have direct evidence of such a structural change for these films. Since the slope change is likely not associated with a change in the Curie temperature, we use the thinner regime ($t<t_C$) to characterize the magnetic properties of the films.

The magnetic dead layer thickness ($t_{DL}$) is estimated from the intercept of the linear curve with the x-axis (Fig. 1(b)). The dead layer thickness is ~0.55 nm for the Ta underlayer films when the thinner regime ($t<t_C$) is used for fitting, whereas it is nearly zero for the thicker range, as reported previously.[2] Note that $\tilde{M}_S$ estimated from the thicker regime is also consistent with previous reports.[2]

The $N_2$ gas concentration dependence of the dead layer thickness is plotted in Fig. 1(c). The dead layer thickness reduces to near zero when Q is more than ~2%. To illustrate the origin of the magnetic dead layer, we show the Q dependence of $t_{DL}$ for the as deposited films in Fig. 1(c), red circles. These results suggest that a dead layer of ~0.3 nm is formed during the deposition process (i.e. before annealing) at the Ta(N)|CoFeB interface for all films, and for films without the nitrogen doping (Q=0), an additional dead layer (~0.3 nm) is formed after annealing, likely by inter-diffusion of Ta and CoFeB. Since TaN is known as a good diffusion barrier,[24] it may help prevent any annealing induced intermixing and perhaps can even sharpen the interface[25] by segregation of the TaN phase[26] (note that $t_{DL}$ decreases upon annealing for Q~9%), although further investigation is required to identify such effect. Recently, it has been reported that a ~0.2 nm thick TaB is formed at the Ta|CoFeB interface,[27] which may be partly responsible for the dead layer formation.

The average saturation magnetization as a function of the nitrogen concentration is shown in Fig. 1(d) for annealed and as deposited films. The Ta underlayer films (Q=0) show the largest $\tilde{M}_S$ after annealing, which is close to that of $Co_{25}Fe_{75}$ as noted by the dashed line in Fig. 1(d). These results show that the Ta underlayer acts as a good boron absorber whereas the nitrogen doped Ta underlayers oppose the absorption process.

The effect of the nitrogen doping significantly influences the underlayer thickness dependence of magnetic moment per unit volume (M/V) and $K_{EFF}$, which are plotted in Figs. 2(a) and 2(b), respectively ($V \equiv A \cdot t$). Here we show results for films annealed at 300 $^{o}$C. For the Ta underlayer, both M/V and $K_{EFF}$ abruptly drop with the underlayer thickness (d) at d~2 nm. In contrast, the nitrogen doped Ta underlayer films exhibit little dependence on d. The strong d dependence of M/V and $K_{EFF}$ for the Ta underlayer films is likely to do with the formation of additional magnetic dead layer after annealing. For studies on the CoFeB layer thickness dependence (Figs. 1 and 3), we fix the underlayer thickness to 1 nm for Ta and 4 nm for TaN(Q>0) underlayers.

To evaluate the magnetic anisotropy of the films, we use the *effective* magnetic layer thickness $t_{EFF} \equiv t - t_{DL}$ hereafter which represents the magnetically active region. The magnetic anisotropy estimated using $t_{EFF}$ is noted as $\tilde{K}_{EFF}$ , which is to be distinguished from $K_{EFF}$ calculated using the nominal CoFeB thickness t. The product of $\tilde{K}_{EFF}$ and $t_{EFF}$ is plotted as a function of $t_{EFF}$ in Fig. 3(a) for the TaN(Q=1%) underlayer films. For all films, $\tilde{K}_{EFF} \cdot t_{EFF}$ shows a linear dependence with $t_{EFF}$ for thicknesses above ~1 nm (and below which M/A vs. t changes its slope), which is in accordance with the expression

$$\tilde{K}_{EFF} \cdot t_{EFF} = \left(K_B - 2\pi M_S{}^2\right) t_{EFF} + K_I \, . \qquad (1)$$

Here $K_B$ and $K_I$ are the bulk and interface contributions to the anisotropy. We fit the data with a linear function to estimate $K_B$ and $K_I$. Positive $K_B$ and $K_I$ represent the easy axis of the corresponding anisotropy directed along the film normal.

The Q dependence of $K_I$ is shown in Fig. 3(b). For the as deposited films, $K_I$ is nearly zero, whereas for the annealed films, surprisingly $K_I$ takes a maximum of ~1.8 erg/cm$^2$ at Q~1%: a ~20% increase from the films with non-doped Ta underlayer. Since the dead layer thickness remains non-zero (~0.3 nm) for Q~1%, it is perhaps more realistic to assume that the nitrogen doped Ta underlayer has improved the PMA of the CoFeB|MgO interface. For example, crystallization of the CoFeB layer, which is assumed to influence the structure of the CoFeB|MgO interface and thus the PMA, may be dependent on the underlayer material that determines the boron diffusion process. Recently, it has been reported[28] that the boron concentration of the CoFeB layer itself can influence $K_I$. From the results shown in Figs. 1(d) and 3(b), we find a similar trend: $K_I$ increases with decreasing boron concentration of the CoFeB layer. With regard to the origin of the enhancement of $K_I$, we note that there is another possibility that the interface between the magnetic dead layer and the CoFeB layer is providing the extra interface PMA.

The slope $\left(K_B - 2\pi M_S^{\,2}\right)$ of the linear fit (Eq. (1)) is plotted against Q in Fig. 3(c). The average saturation magnetization $\tilde{M}_S$, obtained in Fig. 1(d), is substituted into the "slope" expression to extract $K_B$. The Q dependence of $K_B$ is shown in Fig. 3(d). For most of the films, $K_B$ is positive and its magnitude is not negligible. This applies for both the as deposited and the annealed films. Note that for the Ta underlayer films (Q=0), $K_B$ is nearly zero if $t>t_C$ is used to extract $\tilde{M}_S$, which agrees with previous reports[2].

The origin of such a large bulk contribution to the PMA is not clear. Using cross section transmission electron microscopy, we find that the underlayer and the CoFeB layer is predominantly amorphous, although a fraction of the CoFeB layer seems to contain a crystallized region close to the MgO layer.[19, 23, 29] It is difficult to attribute magneto-elastic contribution (i.e. the inverse magnetostriction), if any, to these results since the magnetostriction constants of $Co_{25}Fe_{75}$ and $Co_{20}Fe_{60}B_{20}$ are both positive[30] (assuming that the magnetic layer lattice matches with the MgO layer, a tensile strain will develop in CoFeB which favors in-plane easy axis). We infer that the so-called bond orientation anisotropy (BOA),[31-35] or the single-ion anisotropy, originating from the short range anisotropic alignment of the atomic orbitals, may contribute to developing a large volume PMA. Identification of the origin of the bulk contribution will require a thorough structural and compositional analysis of the film stacks.

In summary, we find that introducing nitrogen into the Ta underlayer increases the interface perpendicular magnetic anisotropy in Ta|CoFeB|MgO heterostructures. The interface anisotropy shows a large dependence on the amount of nitrogen introduced, taking a maximum of ~1.8 erg/cm$^2$ at an optimum nitrogen doping concentration. The magnetic dead layer thickness, likely present at the underlayer|CoFeB interface, monotonically decreases with increasing nitrogen concentration but does not drop to zero for the nitrogen concentration with the maximum interface anisotropy. Our results illustrate the importance of engineering the underlayer|magnetic layer interface to create perpendicularly magnetized ultrathin CoFeB layer, critical for spin transfer and spin-orbit torque devices.

**Acknowledgements**

We thank M. Kodzuka, T. Ohkubo and K. Hono for their support on film characterization. This work was partly supported by the Japan Society for the Promotion of Science (JSPS) through its “Funding program for world-leading innovative R & D on science and technology (FIRST program)”.

## References

[1] A. Brataas, A. D. Kent, and H. Ohno, Nat. Mater. **11**, 372 (2012).
[2] S. Ikeda, K. Miura, H. Yamamoto, K. Mizunuma, H. D. Gan, M. Endo, S. Kanai, J. Hayakawa, F. Matsukura, and H. Ohno, Nat. Mater. **9**, 721 (2010).
[3] D. C. Worledge, G. Hu, D. W. Abraham, J. Z. Sun, P. L. Trouilloud, J. Nowak, S. Brown, M. C. Gaidis, E. J. O'Sullivan, and R. P. Robertazzi, Appl. Phys. Lett. **98**, 022501 (2011).
[4] T. Koyama, et al., Nature Mater. **10**, 194 (2011).
[5] S. W. Jung, W. Kim, T. D. Lee, K. J. Lee, and H. W. Lee, Appl. Phys. Lett. **92**, 202508 (2008).
[6] J. C. Slonczewski, J. Magn. Magn. Mater. **159**, L1 (1996).
[7] L. Berger, Phys. Rev. B **54**, 9353 (1996).
[8] I. M. Miron, et al., Nat. Mater. **10**, 419 (2011).
[9] I. M. Miron, K. Garello, G. Gaudin, P. J. Zermatten, M. V. Costache, S. Auffret, S. Bandiera, B. Rodmacq, A. Schuhl, and P. Gambardella, Nature **476**, 189 (2011).
[10] L. Liu, C.-F. Pai, Y. Li, H. W. Tseng, D. C. Ralph, and R. A. Buhrman, Science **336**, 555 (2012).
[11] P. F. Carcia, J. Appl. Phys. **63**, 5066 (1988).
[12] G. H. O. Daalderop, P. J. Kelly, and F. J. A. Denbroeder, Phys. Rev. Lett. **68**, 682 (1992).
[13] M. T. Johnson, P. J. H. Bloemen, F. J. A. denBroeder, and J. J. deVries, Rep. Prog. Phys. **59**, 1409 (1996).
[14] H. X. Yang, M. Chshiev, B. Dieny, J. H. Lee, A. Manchon, and K. H. Shin, Phys. Rev. B **84**, 054401 (2011).
[15] T. Liu, J. W. Cai, and L. Sun, Aip Advances **2**, 032151 (2012).
[16] S. Ingvarsson, G. Xiao, S. S. P. Parkin, and W. J. Gallagher, J. Magn. Magn. Mater. **251**, 202 (2002).
[17] Y. H. Wang, W. C. Chen, S. Y. Yang, K. H. Shen, C. Park, M. J. Kao, and M. J. Tsai, J. Appl. Phys. **99**, 08M307 (2006).
[18] S. Y. Jang, C. Y. You, S. H. Lim, and S. R. Lee, J. Appl. Phys. **109**, 013901 (2011).
[19] J. Kim, J. Sinha, M. Hayashi, M. Yamanouchi, S. Fukami, T. Suzuki, S. Mitani, and H. Ohno, Nat. Mater. **12**, 240 (2013).
[20] S. S. P. Parkin and M. G. Samant, US Patent 6518588 (2003).
[21] S. M. Rossnagel, J. Vac. Sci. Technol. B **20**, 2328 (2002).
[22] See supplementary information.
[23] S. V. Karthik, Y. K. Takahashi, T. Ohkubo, K. Hono, H. D. Gan, S. Ikeda, and H. Ohno, J. Appl. Phys. **111**, 083922 (2012).
[24] S. M. Rossnagel and H. Kim, J. Vac. Sci. Technol. B **21**, 2250 (2003).
[25] F. J. A. Denbroeder, D. Kuiper, A. P. Vandemosselaer, and W. Hoving, Phys. Rev. Lett. **60**, 2769 (1988).
[26] B. Viala, V. R. Inturi, and J. A. Barnard, J. Appl. Phys. **81**, 4498 (1997).
[27] A. A. Greer, et al., Appl. Phys. Lett. **101**, 202402 (2012).
[28] S. Ikeda, R. Koizumi, H. Sato, M. Yamanouchi, K. Miura, K. Mizunuma, H. D. Gan, F. Matsukura, and H. Ohno, IEEE Trans. Magn. **48**, 3829 (2012).
[29] J. Sinha *et al.*, unpublished.
[30] R. C. Ohandley, Phys. Rev. B **18**, 930 (1978).

[31] Y. Suzuki, J. Haimovich, and T. Egami, Phys. Rev. B **35**, 2162 (1987).
[32] H. Fu and M. Mansuripur, Phys. Rev. B **45**, 7188 (1992).
[33] X. Yan, M. Hirscher, T. Egami, and E. E. Marinero, Phys. Rev. B **43**, 9300 (1991).
[34] S. C. N. Cheng and M. H. Kryder, J. Appl. Phys. **69**, 7202 (1991).
[35] A. T. Hindmarch, A. W. Rushforth, R. P. Campion, C. H. Marrows, and B. L. Gallagher, Phys. Rev. B **83**, 212404 (2011).
[36] R. M. Bozorth, *Ferromagnetism*, Wiley-IEEE Press, New York (1993).

**Figure captions**

Figure 1: (a) Composition ratio of nitrogen to Ta plotted against the $N_2$ sputter gas concentration (Q). The dashed lines indicate the stable $Ta_{1-X}N_X$ compounds (Ref. 21): from bottom, Ta, $Ta_2N$, $TaN_{0.43}$, TaN, $Ta_5N_6$, $Ta_4N_5$ and $Ta_3N_5$. Inset: Q dependence of the single film resistivity ρ. (b) Magnetic moment per unit area (M/A) as a function of the CoFeB layer nominal thickness for $SiO_2$|1 Ta|*t* CoFeB|2 MgO|1 Ta measured at room temperature (circles). Linear fit to the data is shown by the solid line, the dashed line indicates a linear fit for thick (t>2.2 nm) CoFeB films. Squares show M/A measured at 10 K. (c,d) Variation of (c) magnetic dead layer thickness ($t_{DL}$) and (d) average saturation magnetization ($\tilde{M}_S$) with Q for the as deposited (red circles) and 300 $^{o}$C annealed (black squares) films. The dashed and dotted lines in (d) represent reported $M_S$ values of $Co_{25}Fe_{75}$ and $Co_{20}Fe_{60}B_{20}$, respectively (Ref. 36).

Figure 2: (a) Magnetic moment per unit volume (M/V) and (b) magnetic anisotropy ($K_{EFF}$) plotted as a function of the underlayer thickness d for $SiO_2$|d underlayer|1 CoFeB|2 MgO|1 Ta. The underlayer is Ta (black squares), TaN(Q=1%) (red circles) and TaN(Q=9%) (blue triangles). $K_{EFF}$ is estimated using the nominal CoFeB thickness (the dead layer thickness is not subtracted).

Figure 3: Product of the effective anisotropy ($\tilde{K}_{EFF}$) and the effective thickness $t_{EFF}$=t-$t_{DL}$ plotted as a function of $t_{EFF}$. $\tilde{K}_{EFF}$ is estimated using $t_{EFF}$ for the CoFeB thickness. The solid line shows a linear fit to the data for an appropriate data range. (b) Interface magnetic anisotropy ($K_I$), (c) slope of $\tilde{K}_{EFF} \cdot t_{EFF}$ vs. $t_{EFF}$ and (d) bulk contribution to the PMA ($K_B$) as a function of the $N_2$ gas concentration (Q) during the underlayer deposition for the as deposited (red circles) and 300 $^{o}$C annealed (black squares) films. The error bars in (b) correspond to the minimum and maximum values of $K_I$ when the fitting range is varied within the linear part of $\tilde{K}_{EFF} \cdot t_{EFF}$ vs. $t_{EFF}$. Standard error of each fitting is smaller than the size of the symbols. The error bars in (c, d) show the corresponding changes in the slope and $K_B$ when the fitting range is varied. See supplementary information for the details of the fitting.

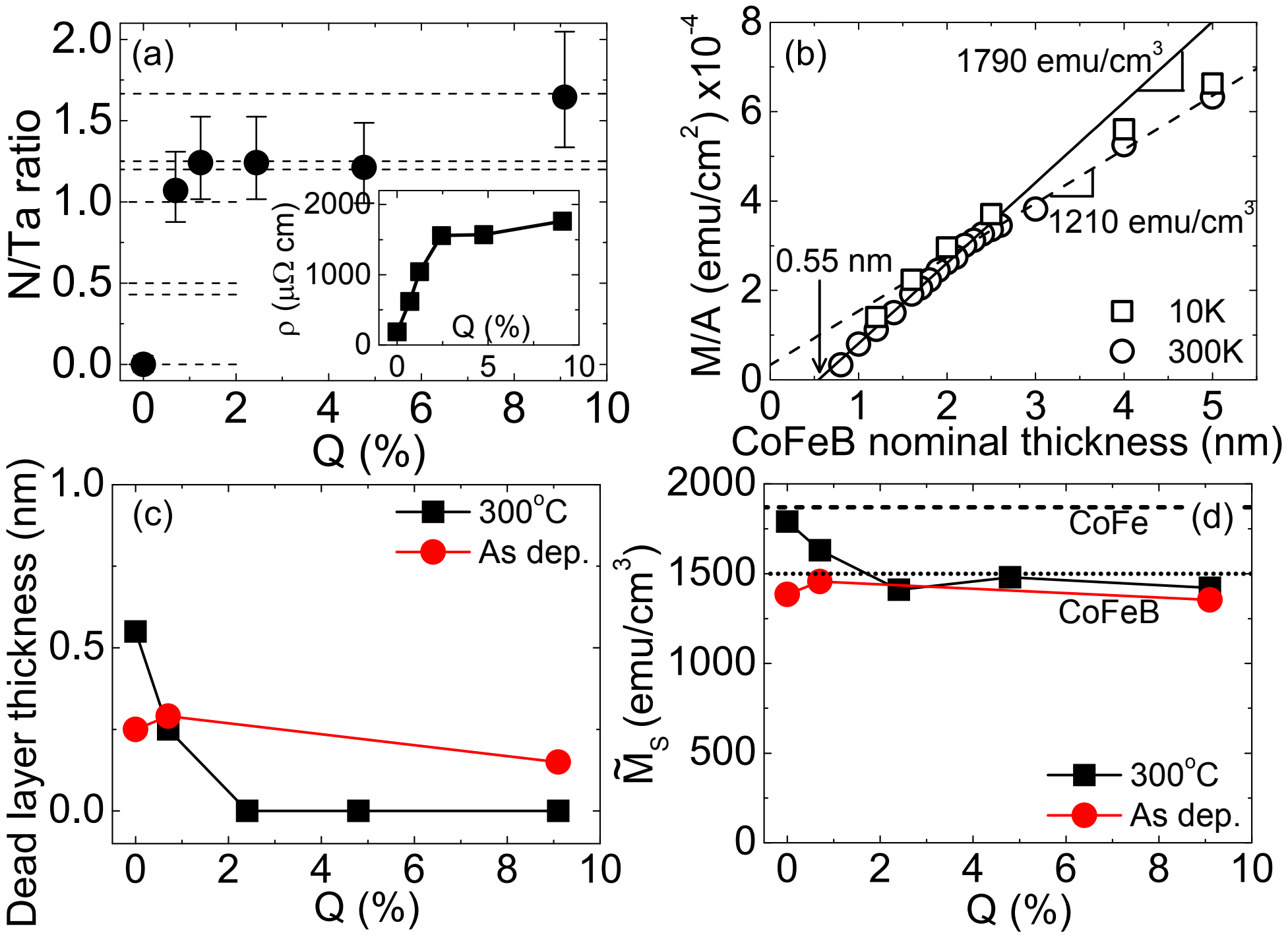

Fig. 1

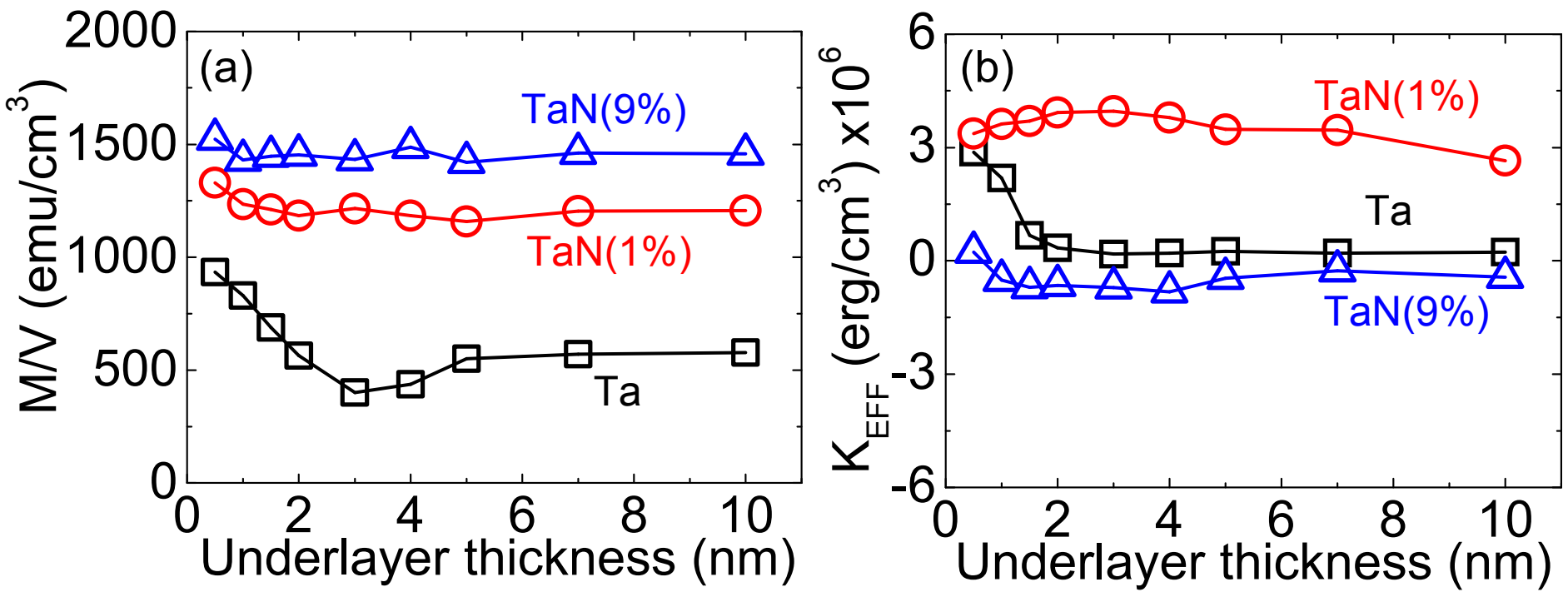

Fig. 2

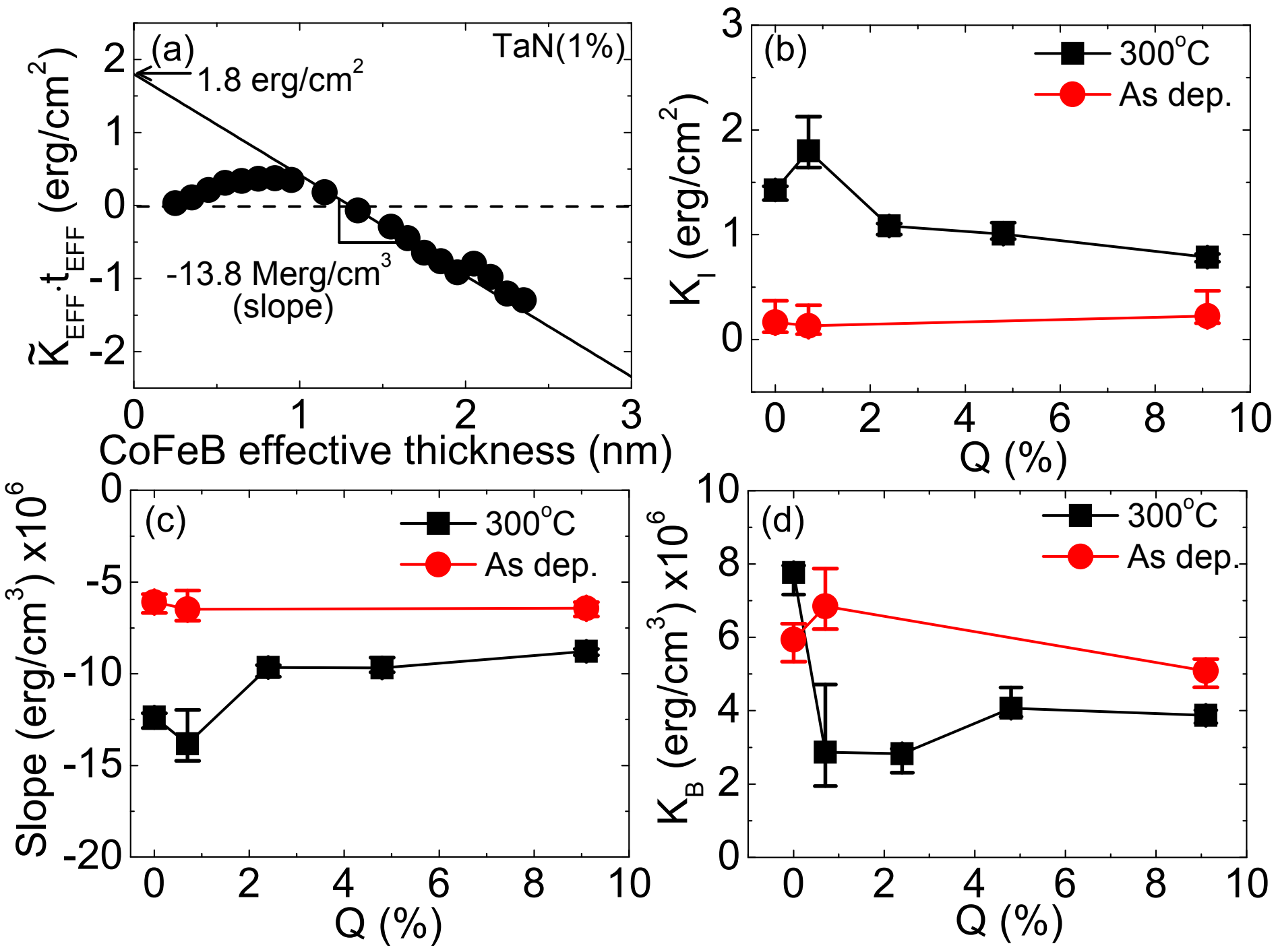

Fig. 3

## Supplementary information for:

# Enhanced interface perpendicular magnetic anisotropy in Ta|CoFeB|MgO using nitrogen doped Ta underlayers

Jaivardhan Sinha[1], Masamitsu Hayashi[1*], Andrew J. Kellock[2], Shunsuke Fukami[3], Michihiko Yamanouchi[3,4], Hideo Sato[3], Shoji Ikeda[3,4], Seiji Mitani[1], See-hun Yang[2], Stuart S. P. Parkin[2] and Hideo Ohno[3,4,5]

[1]*National Institute for Materials Science, Tsukuba 305-0047, Japan*

[2]*IBM Almaden Research Center, San Jose, CA 95120, USA*

[3] *Center for Spintronics Integrated Systems, Tohoku University, Sendai 980-8577, Japan*

[4]*Laboratory for Nanoelectronics and Spintronics, Research Institute of Electrical Communication, Tohoku University, Sendai 980-8577, Japan*

[5]*WPI Advanced Institute for Materials Research, Tohoku University, Sendai 980-8577, Japan*

**Figure captions**

Figure S1: Composition ratio of nitrogen to Ta versus resistivity for the single layer films. $N_2$ concentration (Q) during the underlayer deposition is inserted next to each point. The stable compounds of nitrogen doped Ta are shown by the horizontal dashed lines.

Figure S2: Magnetic moment per unit area vs. nominal CoFeB layer thickness. Linear fitting is used to extract the magnetic dead layer thickness ($t_{DL}$) and the saturation magnetization ( $\tilde{M}_S$ ). Left columns: data for 300˚C annealed films. Right columns: data for as deposited films. All films have the following film structure: $SiO_2$|d underlayer|t CoFeB|2 MgO|1 Ta. The underlayer is (a,b) 1 Ta, (c,d) 4 TaN (Q=1%) (e) 4 TaN(Q=2%), (f) 4 TaN(Q=5%) and (g,h) 4 TaN(Q=9%). The extracted dead layer thickness and the saturation magnetization are shown in each panel. For some of the film stacks, M/A deviates from the linear fit for CoFeB thicknesses below ~0.6 nm: this may be due to superparamagnetism if the CoFeB layer forms an island-like structure at the initial stage of the growth process (i.e. for thin films).

Figure S3: Product of the effective magnetic anisotropy ( $\tilde{K}_{EFF}$ ) and the effective thickness ($t_{EFF}$ =t-$t_{DL}$) is plotted against $t_{EFF}$. Linear fitting is used to extract the interface ($K_I$) and bulk ($K_B$-2π$M_S^2$) contribution to the magnetic anisotropy. Since $K_I$ and $K_B$ depends on the fitting range, we have shown three representative linear fittings by the solid, dashed and dash-dotted lines. The blue dash-dotted and red dashed lines indicate the range where the maximum and minimum $K_I$ are found within the linear part of $\tilde{K}_{EFF} \cdot t_{EFF}$ vs. $t_{EFF}$, respectively. The black solid line represents the fitting range used to obtain the value indicated by the symbols in Fig. 3(b). The corresponding fitting range is indicated by the horizontal colored arrows in each panel. Left columns: data for 300˚C annealed films. Right columns: data for as deposited films. All films have the following film structure: $SiO_2$|d underlayer|t CoFeB|2 MgO|1 Ta. The underlayer is (a,b) 1 Ta, (c,d) 4 TaN (Q=1%) (e) 4 TaN(Q=2%), (f) 4 TaN(Q=5%) and (g,h) 4 TaN(Q=9%). The extracted values of $K_I$ and $K_B$-2π$M_S^2$ are shown in each panel.

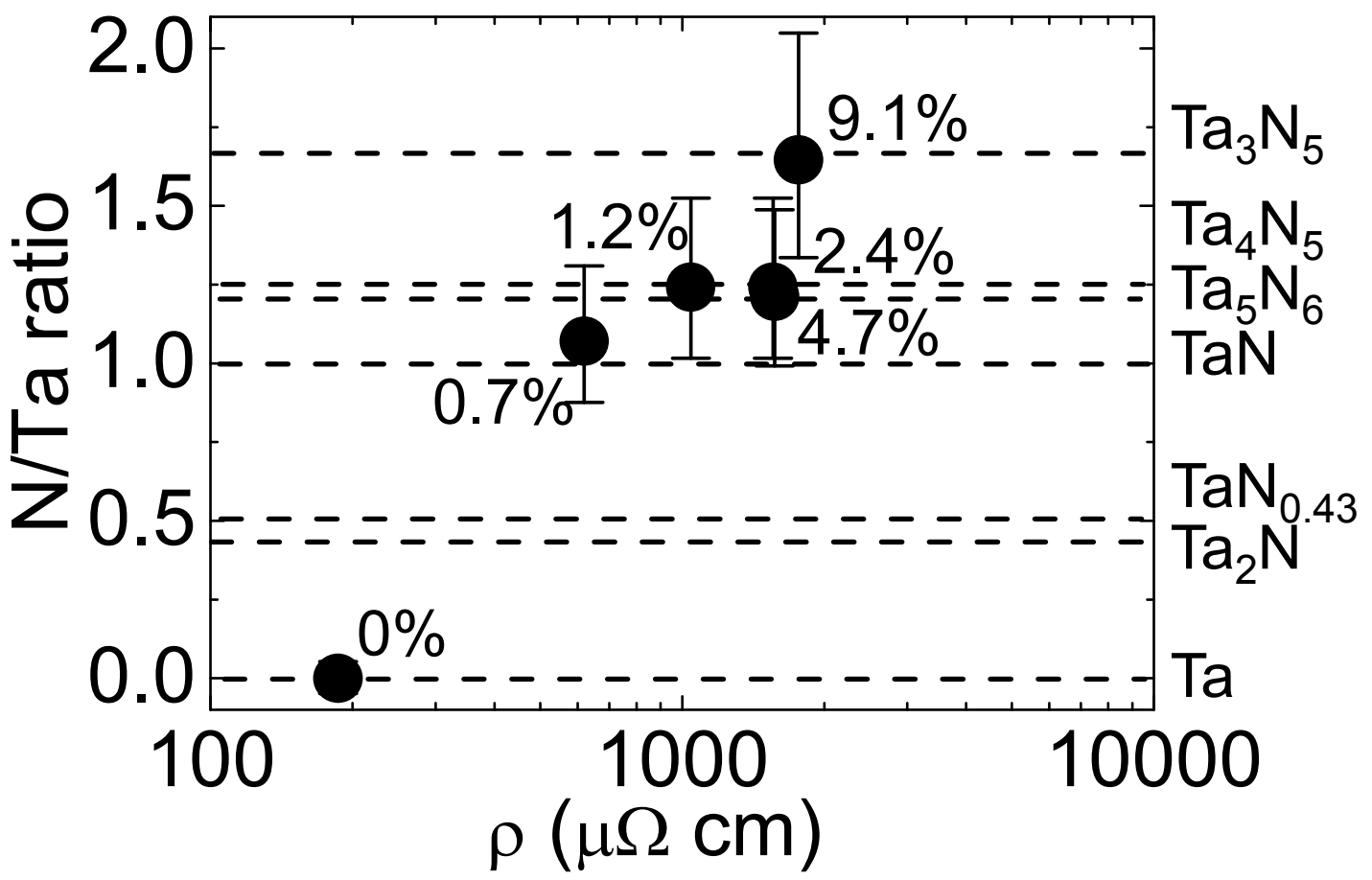

Fig. S1

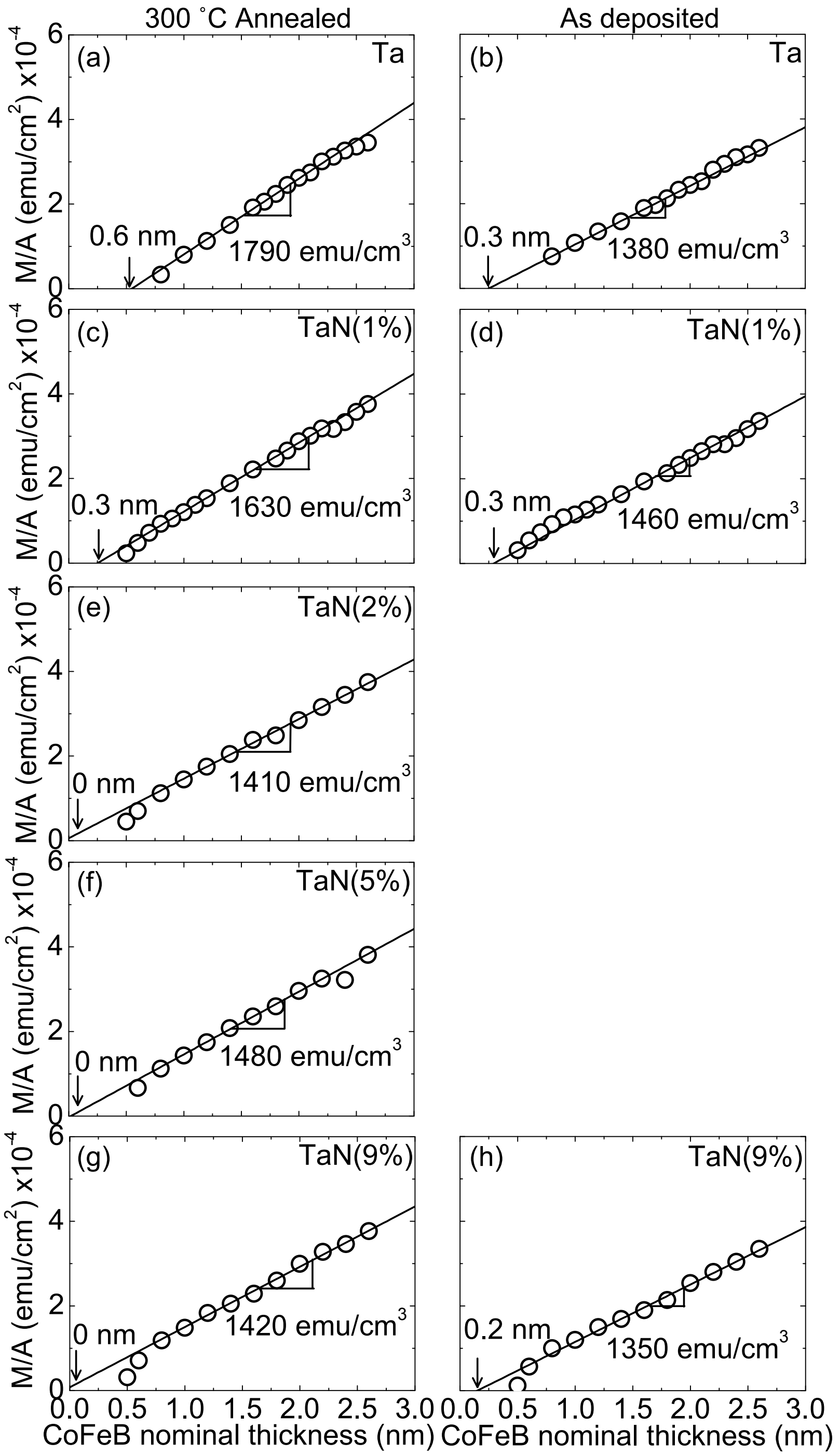

Fig. S2

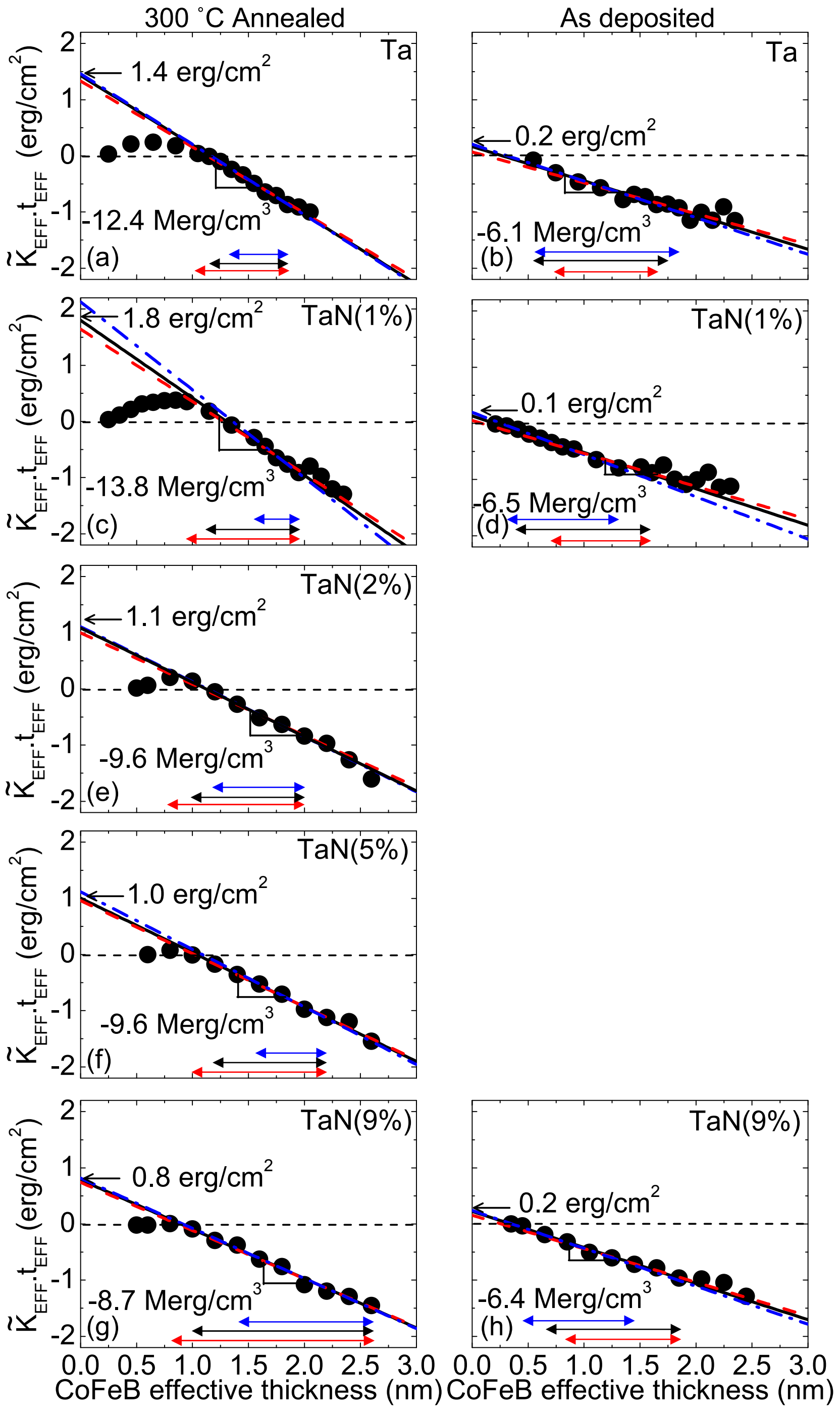

Fig. S3